\documentclass[pra, twosides, twocolumn]{revtex4}
\usepackage{graphicx, amsmath, amssymb}

\begin{document}

\title{Decoherence Free Subspace and entanglement by interaction with a common
squeezed bath.}
\author{D. Mundarain$^{1}$ and M. Orszag$^{2}$}
\address{ \
${}^{1}$ Departments de  F\'{\i}sica, Universidad Simon Bolívar,
Apartado Postal 89000, Caracas 1080A, Venezuela, \ ${}^{2}$
Faculad de F\'{\i}sica, Pontificia Universidad Cat\'{o}lica de
Chile, Casilla 306, Santiago, Chile}

\begin{abstract}
In this work we find explicitly the decoherence free subspace (DFS) for a
two two-level system in a common squeezed vacuum bath. We also find an
orthogonal basis for the DFS composed of a symmetrical and an
antisymmetrical (under particle permutation) entangled state. For any
initial symmetrical state, the master equation has one stationary state
which is the symmetrical entangled decoherence free state. In this way, one
can generate entanglement via common squeezed bath of the two systems. If
the initial state does not have a definite parity, the stationary state
depends strongly on the initial conditions of the system and it is a
statistical mixture of states which belong to DFS. We also study the effect
of the coupling between the two-level systems on the DFS.
\end{abstract}

\maketitle

One of the most important challenges of Quantum Computation is to revert the
effects of the environment over systems used to store information. In the
general case, these interactions limit the capability of a quantum computer
\cite{gruska}, \cite{yu}, \cite{nielsen}, \cite{giulini}. There are some
interesting proposal to circumvent the injurious influence of reservoirs,
one of them is related to the use of Decoherence Free Subspace (DFS) as the
memory space for storing the quantum information. The search of ways to
bypass decoherence in Quantum Information Processing (QIP) started with
Palma \textit{et al }\cite{palma}, in a study of the dephasing of two qubits
in contact with a reservoir. Duan and Gao \cite{duan}, used this model with
a different method and coined the term ''Coherence preserving states''. The
general framework for DFS was done by Zanardi \textit{et al} \cite{za} in a
spin-boson model, in a ''collective decoherence mode'', undergoing both
dephasing and dissipation. Based on dynamic symmetries in the
spin-environment interaction, they provide a general condition for the
decoherence free states.

Also, Lidar \textit{et e}l \cite{li1} introduced the term ''decoherence-free
subspaces'' (DFS) and analyzed the robustness of these states against
perturbations. A completely general condition for the existence of the DF
states in terms of the Kraus operators was provided by Lidar, Bacon and
Whaley \cite{li2}. A simple definition of the DFS, is the following one:
\textit{a system with Hilbert space }$H$\textit{\ is said to have a
decoherence free subspace }$\widetilde{H}$\textit{\ if the evolution inside }%
$\widetilde{H}\subset H$\textit{\ is purely unitary.}

In the presence of the environment, the DFS is a set of all states which are
not affected at all by the interaction with the bath. In terms of the
reduced dynamics of the system, they are invariant states. Since not all
quantum open systems have a DFS , it is an important issue to study the
systems which do and also investigate the dynamical properties of such
systems.\cite{ba}, \cite{be1}, \cite{be2}, \cite{viola}

In this letter we are concerned with a two two-level system interacting with
a common squeezed vacuum bath. We show that this system has a DFS. We will
review some properties of the stationary states of this system and will
establish the relations between them and the DFS. We will also address
another important issue: the preparation of entangled states. We will show
that for any initial symmetrical state, state which is invariant under
particle permutation, the stationary state depends on the squeezing
parameters of the bath, and most importantly, it is pure and entangled.

The problem of the stationary states of a master equation, or expressed more
mathematically, the stationary states of quantum dynamical semigroups is, in
general , an involved problem.\cite{fri} In our case, we derive directly
from the master equation a set of linear differential equations which are
solved explicitly in order to obtain the stationary states of the system.
The stationary states depend strongly on the initial conditions and they are
statistical mixtures of states which belong to the DFS.

Lets first consider the master equation, in the interaction picture, for a
two level system in a broadband squeezed vacuum \cite{ga}:
\begin{eqnarray}
\frac{\partial \rho }{\partial t} &=&\frac{1}{2}\gamma \left( N+1\right)
\left( 2\sigma \rho \sigma ^{\dagger }-\sigma ^{\dagger }\sigma \rho -\rho
\sigma ^{\dagger }\sigma \right)  \nonumber  \label{em1} \\
&&+\frac{1}{2}\gamma N\left( 2\sigma ^{\dagger }{\rho }\sigma -\sigma \sigma
^{\dagger }{\rho }-{\rho }\sigma \sigma ^{\dagger }\right)  \nonumber \\
&&-\frac{1}{2}\gamma Me^{i\psi }\left( 2\sigma ^{\dagger }{\rho }\sigma
^{\dagger }-\sigma ^{\dagger }\sigma ^{\dagger }{\rho }-{\rho }\sigma
^{\dagger }\sigma ^{\dagger }\right)  \nonumber \\
&&-\frac{1}{2}\gamma Me^{-i\psi }\left( 2\sigma {\rho }\sigma -\sigma \sigma
{\rho }-{\rho }\sigma \sigma \right)
\end{eqnarray}
where $\gamma $ is the vacuum decay constant and $N,M=\sqrt{N(N+1)}$ and $%
\psi $ are the squeeze parameters of the bath. The two Pauli spin matrices
are:

\begin{equation}
\sigma ^{\dagger }=\left(
\begin{array}{cc}
0 & 1 \\
0 & 0
\end{array}
\right) \qquad \sigma =\left(
\begin{array}{cc}
0 & 0 \\
1 & 0
\end{array}
\right)  \label{l2}
\end{equation}

It is not difficult to show that the master equation can be written in an
explicit Lindblad form using only one Lindblad operator:
\begin{equation}
\frac{\partial \rho }{\partial t}=\frac{\gamma }{2}\left\{ 2S\rho S^{\dagger
}-\rho S^{\dagger }S-S^{\dagger }S\rho \right\}  \label{em2}
\end{equation}
where
\begin{eqnarray}
S &=&\sqrt{N+1}\sigma -\sqrt{N}\exp \left\{ i\psi \right\} \sigma ^{\dagger }
\label{l3} \\
&=&\cosh (r)\sigma -\sinh (r)\exp \left\{ i\psi \right\} \sigma ^{\dagger }
\end{eqnarray}
This operator has two eigenstates:
\begin{equation}
S|\lambda _{\pm }\rangle =\lambda _{\pm }|\lambda _{\pm }\rangle  \label{l4}
\end{equation}
where
\begin{equation}
|\lambda _{\pm }\rangle =\sqrt{\frac{N}{N+M}}|+\rangle \mp i\sqrt{\frac{M}{%
N+M}}e^{-i\psi /2}|-\rangle  \label{l5}
\end{equation}
with eigenvalues $\lambda _{\pm }=\pm i\sqrt{M}\exp \{i\psi /2\}$.

For a two two-level system interacting with a common squeezed bath, the
master equation has the same structure as in the one particle case, but now:
\begin{equation}
S=\sqrt{N+1}\Sigma -\sqrt{N}\exp \left\{ i\psi \right\} \Sigma ^{\dagger }
\label{l6}
\end{equation}
where the $\Sigma $ 's are the combined ladder operators for the two
particles:
\begin{equation}
\Sigma =\sigma _{1}+\sigma _{2},\quad \Sigma ^{\dagger }=\sigma
_{1}^{\dagger }+\sigma _{2}^{\dagger }  \label{l7}
\end{equation}
Thus, the two particle Lindblad operator can be written as the sum of the
two Lindblad operators of each particle:
\begin{equation}
S=S_{1}+S_{2}  \label{l8}
\end{equation}
In this case the DFS \cite{li2} is composed of all eigenstates of $S$ with
zero eigenvalues. The following two linearly independent states satisfy this
property.

\begin{equation}
|\psi _{1}\rangle =|\lambda _{+}\rangle _{1}\otimes \ |\lambda _{-}\rangle
_{2}  \label{l9}
\end{equation}
\begin{equation}
|\psi _{2}\rangle =|\lambda _{-}\rangle _{1}\otimes \ |\lambda _{+}\rangle
_{2}  \label{l10}
\end{equation}
where the subscripts 1,2 refer to the particle one and two respectively.
These two states are not orthogonal. They define a plane in the Hilbert
space, spanned by the following orthonormal states:

\begin{equation}
|\phi _{1}\rangle =\frac{1}{\sqrt{N^{2}+M^{2}}}\left( N|+,+\rangle
+Me^{-i\psi }|-,-\rangle \right) ,  \label{l11}
\end{equation}
\begin{equation}
|\phi _{2}\rangle =\frac{1}{\sqrt{2}}\left( |-,+\rangle -|+,-\rangle \right)
,  \label{l12}
\end{equation}
where $|\pm ,\pm \rangle =|\pm \rangle _{1}\otimes \ |\pm \rangle _{2}$ and $%
|\pm ,\mp \rangle =|\pm \rangle _{1}\otimes \ |\mp \rangle _{2}$.

Now, one can define two other states which are orthogonal to $|\phi
_{1}\rangle $ and $|\phi _{2}\rangle $:

\begin{equation}
|\phi _{3}\rangle =\frac{1}{\sqrt{2}}\left( |-,+\rangle +|+,-\rangle \right)
,  \label{l13}
\end{equation}
\begin{equation}
|\phi _{4}\rangle =\frac{1}{\sqrt{N^{2}+M^{2}}}\left( M|+,+\rangle
-Ne^{-i\psi }|-,-\rangle \right) .  \label{l14}
\end{equation}

In the basis $\{|\phi _{1}\rangle ,|\phi _{2}\rangle ,|\phi _{3}\rangle
,|\phi _{4}\rangle \}$ the Lindblad operator $S$ for the two particles has a
very simple form:
\begin{equation}
s=\left(
\begin{array}{cccc}
0 & 0 & \alpha e^{i\psi } & 0 \\
0 & 0 & 0 & 0 \\
0 & 0 & 0 & \delta e^{i\psi } \\
0 & 0 & \beta & 0
\end{array}
\right)  \label{l15}
\end{equation}
where
\begin{equation}
\alpha =\sqrt{\frac{2}{2N+1}},  \label{l16}
\end{equation}
\begin{equation}
\beta =-2\sqrt{N(N+1)}\alpha ,  \label{17}
\end{equation}
\begin{equation}
\delta =\frac{2}{\alpha }.  \label{l18}
\end{equation}
In this basis the master equation becomes a system of fifteen differential
equations. The various components of the system are either constant or
exponentially decaying terms with rates that depend on $\alpha $, $\beta $
and $\delta $. The exponentially decaying terms go to zero and eventually
one finds the following stationary state for the master equation:

\begin{equation}
\rho (\infty )=\rho _{ss}=\left(
\begin{array}{cccc}
1-\rho _{22}(0) & \rho _{12}(0) & 0 & 0 \\
\rho _{21}(0) & \rho _{22}(0) & 0 & 0 \\
0 & 0 & 0 & 0 \\
0 & 0 & 0 & 0
\end{array}
\right)  \label{l19}
\end{equation}
The first important thing to notice from this equation is that when the
initial state does not contain the antisymmetrical state $|\phi _{2}\rangle $
($\rho _{22}(0)=\rho _{12}(0)=\rho _{21}(0)=0$) the stationary state of the
master equation is the pure entangled state $|\phi _{1}\rangle $:
\begin{equation}
\rho _{ss}=|\phi _{1}\rangle \langle \phi _{1}|=\left(
\begin{array}{cccc}
1 & 0 & 0 & 0 \\
0 & 0 & 0 & 0 \\
0 & 0 & 0 & 0 \\
0 & 0 & 0 & 0
\end{array}
\right) .  \label{l20}
\end{equation}

In figure (1) we show the time evolution of the probability $\langle \phi
_{1}|\rho (t)|\phi _{1}\rangle $ for the initial symmetrical state $%
|++\rangle $. As one can see this quantity goes to 1 as $t\rightarrow \infty
$. This fact indicates that the final state is the decoherence free state $%
|\phi _{1}\rangle $. This result also shows that one can generate
entanglement via a common squeezed bath of the two two-level systems.
\begin{figure}[h]
\includegraphics[width=6cm,angle=-90]{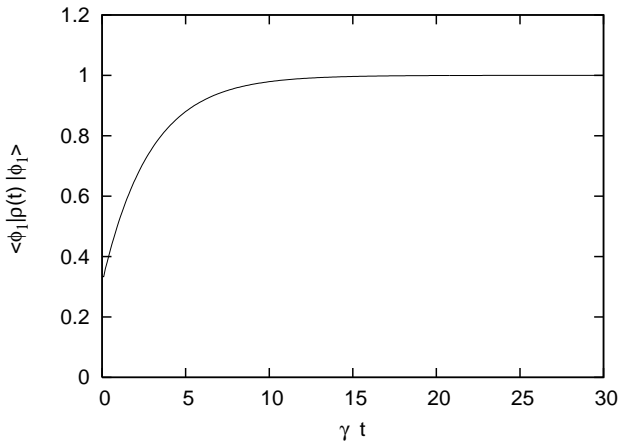}\bigskip
\caption{Probability for the system to be in $|\phi _{1}\rangle $ as a
function of time. We took $N=1$, $\psi =0,$ and the initial state is $%
|++\rangle $. One starts with a factorized state and generates an entangled
one via common squeezed bath.}
\label{etiqueta1}
\end{figure}

In any other case, the stationary states depends on the initial states of
the system, but in general it is a mixed of state \textit{that belongs to
the DFS of the master equation}.

\begin{equation}
\rho _{ss}=P_{1}|\nu _{1}\rangle \langle \nu _{1}|+P_{2}|\nu _{2}\rangle
\langle \nu _{2}|  \label{l21}
\end{equation}
where
\begin{equation}
|\nu _{1}\rangle =A_{1}\left( (P_{1}-\rho _{22}(0))|\phi _{1}\rangle +\rho
_{21}(0)|\phi _{2}\rangle \right)  \label{l22}
\end{equation}
\begin{equation}
|\nu _{2}\rangle =A_{2}\left( (P_{2}-\rho _{22}(0))|\phi _{1}\rangle +\rho
_{21}(0)|\phi _{2}\rangle \right)  \label{l23}
\end{equation}
where the normalization constants $A_{1},A_{2}$ are
\begin{equation}
A_{1,2}=\frac{1}{\sqrt{\mid \rho _{12}(0)\mid ^{2}-(P_{1,2}-\rho
_{22}(0))^{2}}}  \label{l24}
\end{equation}
and
\begin{equation}
P_{1,2}=\frac{1}{2}\pm \sqrt{\left( \frac{1}{2}\right) ^{2}+\mid \rho
_{12}(0)\mid ^{2}-\rho _{22}(0)\left( 1-\rho _{22}(0)\right) }  \label{l25}
\end{equation}
One can quantify the purity of the final states observing that as
\begin{equation}
\mathrm{Tr}\left( \rho _{ss}{}^{2}\right) \leq 1  \label{l26}
\end{equation}
then
\begin{equation}
\mid \rho _{12}(0)\mid ^{2}\leq \rho _{22}(0)(1-\rho _{22}(0))  \label{l27}
\end{equation}
From this expression we define a parameter $x$ which characterizes the
initial coherence between the states $|\phi _{1}\rangle $ and $|\phi
_{2}\rangle $:
\begin{equation}
\mid \rho _{12}(0)\mid ^{2}=x\rho _{22}(0)(1-\rho _{22}(0)).  \label{l28}
\end{equation}
From the initial considerations this parameter also characterize the purity
of final state. From this we obtain three cases in which the final states
are pure:

\begin{enumerate}
\item  $\rho _{22}(0)=0$, which corresponds to the case of having a
symmetrical initial state; the stationary state is $|\phi _{1}\rangle $ .

\item  $\rho _{22}(0)=1$ which corresponds to the case of having $|\phi
_{2}\rangle $ as initial state; this state does not evolve because it
belongs to the DFS. It is an invariant state.

\item  $\rho _{22}(0)\neq 0\,\mathrm{or}\,1$ and $x=1$, in this case the
initial state must be a pure state which is a linear combination of $|\phi
_{1}\rangle $ and $|\phi _{2}\rangle $. This state is also an invariant
state.
\end{enumerate}

In all other cases, the stationary state is a mixed state, but it remains
entangled.

We consider the interesting limit $N\rightarrow 0$ (vacuum bath). In this
limit, the states $|\phi _{1}\rangle $ and $|\phi _{4}\rangle $ of the
orthonormal basis become:
\begin{equation}
|\phi _{1}\rangle =|--\rangle \quad \mathrm{and}\quad |\phi _{4}\rangle
=|++\rangle  \label{l29}
\end{equation}
and the other two states of the basis do not change. The stationary state
has the same previous structure with the new basis. Any symmetrical state
decays to $|\phi _{1}\rangle =|--\rangle $. In particular
\begin{equation}
|++\rangle \rightarrow |--\rangle  \label{l30}
\end{equation}
It is interesting to observe that when the initial state is not completely
symmetrical, the system does not decay to the $|--\rangle $ state. For
example, for the initial condition:
\begin{equation}
|-+\rangle =\frac{1}{\sqrt{2}}\left( |\phi _{2}\rangle +|\phi _{3}\rangle
\right)  \label{l31}
\end{equation}
one has:

\begin{equation}
\rho _{22}(0)=\frac{1}{2}\quad \mathrm{and}\quad \rho _{12}(0)=0  \label{l32}
\end{equation}
and we get the following result:
\begin{equation}
|-+\rangle \rightarrow \frac{1}{2}|\phi _{1}\rangle \langle \phi _{1}|+\frac{%
1}{2}|\phi _{2}\rangle \langle \phi _{2}|  \label{l33}
\end{equation}

For a higher dimensional problem, for instance, when the number of spins $%
\mathcal{N=}4$ we can form the products
\begin{equation}
|\psi _{1}\rangle =|\lambda _{+}\rangle _{1}\otimes \ |\lambda _{+}\rangle
_{2}\otimes |\lambda _{-}\rangle _{3}\otimes \ |\lambda _{-}\rangle _{4},
\end{equation}

and cyclic permutations (6 possible combinations), which belong to the DFS
since we have the same number of equal positive and negative eigenvalues of
S, adding up to zero.

For even $\mathcal{N}$, the dimension of the DFS is
\begin{equation}
DIM(DFS\text{ of }\mathcal{N}\text{ spins})=\frac{\mathcal{N}\text{!}}{%
\left[ (\frac{\mathcal{N}}{2})!\right] ^{2}}.
\end{equation}

In general, in order to have 2 atoms in a common bath, they will have to be
quite near, at a distance no bigger than the correlation length of the bath.
This implies that one cannot avoid an interaction between the particles.
This interaction between the two-level systems can, in principle, affect the
DFS. For example, one can consider a Dipole-Dipole Van der Waals coupling of
the form:
\begin{equation}
H_{D}=\hbar \Omega (\sigma _{1}\sigma _{2}^{\dagger }+\sigma _{1}^{\dagger
}\sigma _{2}),
\end{equation}
with $\Omega =\mid \mathbf{d}\mid ^{2}\frac{(1-3\cos ^{2}\theta )}{R^{3}}$,
where R is the modulus of the distance between the atoms and $\theta $ the
angle between the separation vector and $\mathbf{d}$ (dipole matrix).

It is interesting to study the effect of such a Hamiltonian over our
decoherence free states $|\phi _{1}\rangle $ and $|\phi _{2}\rangle $. It is
simple to verify that:
\begin{eqnarray}
(\sigma _{1}\sigma _{2}^{\dagger }+\sigma _{1}^{\dagger }\sigma _{2})|\phi
_{1}\rangle  &=&0, \\
(\sigma _{1}\sigma _{2}^{\dagger }+\sigma _{1}^{\dagger }\sigma _{2})|\phi
_{2}\rangle  &=&-|\phi _{2}\rangle .  \nonumber
\end{eqnarray}
As we can see, for an initial state within the DFS, with this type of
coupling the state remains within the DFS. As a matter of fact, the time
evolution operator just introduces a time dependent phase factor in $|\phi
_{2}\rangle $ and leaves $|\phi _{1}\rangle $ invariant:
\begin{eqnarray}
|\phi _{2}\rangle  &\rightarrow &\exp \left[ -i\Omega (\sigma _{1}\sigma
_{2}^{\dagger }+\sigma _{1}^{\dagger }\sigma _{2})t\right] |\phi _{2}\rangle
=\exp \left[ i\Omega t\right] |\phi _{2}\rangle ,  \nonumber \\
|\phi _{1}\rangle  &\rightarrow &\exp \left[ -i\Omega (\sigma _{1}\sigma
_{2}^{\dagger }+\sigma _{1}^{\dagger }\sigma _{2})t\right] |\phi _{1}\rangle
=|\phi _{1}\rangle .
\end{eqnarray}
Thus, the dipole-dipole coupling does not affect the DFS of two two-level
systems.

The Ising- type Hamiltonian $H=A\sum_{i=1}^{\mathcal{N}}S_{i}^{z}S_{i+1}^{z}$
for $\mathcal{N}$ two-level systems is another example. It has , again the
following effect: if one starts with a mixed state within the DFS, and since
$S^{z}|\lambda _{\pm }\rangle =|\lambda _{\mp }\rangle $, the system will
remain in the DFS.

To summarize, we have found a DFS for two two-level system in a common
squeezed bath, and found the relation between this subspace and the steady
state for any initial condition. In some particular cases, when the initial
condition is symmetrical, we get a steady state that is pure and entangled.
However, in the most general case, the steady state is in a mixed state
within the DFS. This is an interesting property of this particular system,
but it is not  necessarily true for other systems.

Finally, we calculate the dimension of the general DFS for $\mathcal{N}$
two-level systems, and we also discuss the anavoidable effect of the
coupling between these two-level systems.

\subsection{Acknowledgements}

D.M. was supported by Did-Usb Grant Gid-30 and by Fonacit Grant No
G-2001000712.

M.O was supported by Fondecyt \# 1051062 and Nucleo Milenio ICM(P02-049)

The authors thank Prof. Sascha Wallentowitz for useful discussions.

\end{document}